\documentclass[10pt, a4paper, twocolumn, copyright, goog]{google}

\usepackage{microtype}
\usepackage{tikz}
\usetikzlibrary{positioning, arrows.meta}

\usepackage[backend=biber, style=authoryear, sortcites=true, natbib=true]{biblatex}
\DeclareSourcemap{
  \maps[datatype=bibtex]{
    \map{
      \step[fieldset=archivePrefix,fieldvalue=arxiv]
    }
  }
}
\DeclareFieldFormat{doi}{%
  \ifhyperref
    {\href{https://doi.org/#1}{doi:#1}}
    {doi:#1}}

\DeclareFieldFormat{eprint:arxiv}{%
  \ifhyperref
    {\href{https://arxiv.org/abs/#1}{arXiv:#1}}
    {arXiv:#1}}

\AtEveryBibitem{%
  \ifboolexpr{
    not test {\iffieldundef{doi}}
    or
    not test {\iffieldundef{eprint}}
  }
    {\clearfield{url}}
    {}%
}
\usepackage{xcolor}

\usepackage{csquotes}
\usepackage[acronym]{glossaries}
\makeglossaries

\definecolor{abbrevblue}{HTML}{000066}

\newacronym{ai}{AI}{Artificial Intelligence}
\newacronym{qc}{QC}{Quantum Computing}
\newacronym{qml}{QML}{Quantum Machine Learning}
\newacronym{ml}{ML}{Machine Learning}
\newacronym{qec}{QEC}{Quantum Error Correction}
\newacronym{nisq}{NISQ}{Noisy Intermediate-Scale Quantum}
\newacronym{vqe}{VQE}{Variational Quantum Eigensolver}
\newacronym{qaoa}{QAOA}{Quantum Approximate Optimization Algorithm}
\newacronym{qram}{QRAM}{Quantum Random Access Memory}
\newacronym{qrom}{QROM}{Quantum Read-Only Memory}
\newacronym{ldpc}{LDPC}{Low-Density Parity-Check}
\newacronym{qldpc}{qLDPC}{Quantum Low-Density Parity-Check}
\newacronym{vqa}{VQA}{Variational Quantum Algorithm}
\newacronym{vqc}{VQC}{Variational Quantum Circuit}
\newacronym{llm}{LLM}{Large Language Model}
\newacronym{qnn}{QNN}{Quantum Neural Network}
\newacronym{qcnn}{QCNN}{Quantum Convolutional Neural Network}
\newacronym{qgnn}{QGNN}{Quantum Graph Neural Network}
\newacronym{pqc}{PQC}{Parameterized Quantum Circuit}
\newacronym{dqc}{DQC}{Differentiable Quantum Circuit}
\newacronym{vsql}{VSQL}{Variational Shadow Quantum Learning}
\newacronym{qcbm}{QCBM}{Quantum Circuit Born Machine}
\newacronym{qrl}{QRL}{Quantum Reinforcement Learning}
\newacronym{dqn}{DQN}{Deep Q-Network}
\newacronym{vqls}{VQLS}{Variational Quantum Linear Solver}
\newacronym{qkm}{QKM}{Quantum Kernel Method}
\newacronym{svm}{SVM}{Support Vector Machine}
\newacronym{pqk}{PQK}{Projected Quantum Kernel}
\newacronym{qla}{QLA}{Quantum Linear Algebra}
\newacronym{hhl}{HHL}{Harrow-Hassidim-Lloyd}
\newacronym{qls}{QLS}{Quantum Linear System}
\newacronym{qsvd}{QSVD}{Quantum Singular Value Decomposition}
\newacronym{qpca}{QPCA}{Quantum Principal Component Analysis}
\newacronym{qa}{QA}{Quantum Annealing}
\newacronym{qubo}{QUBO}{Quadratic Unconstrained Binary Optimization}
\newacronym{dqi}{DQI}{Decoded Quantum Interferometry}
\newacronym{dft}{DFT}{Density Functional Theory}
\newacronym{xc}{XC}{Exchange-Correlation}
\newacronym{pes}{PES}{Potential Energy Surface}
\newacronym{rbm}{RBM}{Restricted Boltzmann Machine}
\newacronym{pinn}{PINN}{Physics-Informed Neural Network}
\newacronym{nqs}{NQS}{Neural Quantum State}
\newacronym{tls}{TLS}{Two-Level System}
\newacronym{rl}{RL}{Reinforcement Learning}
\newacronym{qas}{QAS}{Quantum Architecture Search}

\DeclareFieldFormat{note}{\par\addvspace{0.5em}{\color{blue}\small#1}}

\newcommand{\mytocdepth}{1}


\uselogo{}

\title{The Virtuous Cycle of Quantum-Classical Machine Learning}

\correspondingauthor{nenadt@google.com}

\author[1]{Nenad Toma\v{s}ev} \author[2]{Jarrod R. McClean} \author[1]{Johannes Bausch}

\affil[1]{Google DeepMind}
\affil[2]{Google Quantum AI}

\begin{abstract}
Artificial intelligence and quantum computing have both seen tremendous progress in recent years, opening up new avenues for accelerating scientific discovery.
Classical machine learning has already proven to be useful for addressing challenges in quantum computation; and hardware progress is well underway towards the early fault-tolerant regime.
On the other hand, the emerging field of quantum machine learning is aimed at utilizing the strengths of both computational paradigms, via the development of learning algorithms or models that benefit from running on quantum devices or from training on quantum data.
We therefore argue that both of these paradigms stand to benefit substantially from each other.
Here we review the most salient opportunities for machine learning in quantum computing, hoping to inspire a virtuous cycle through which both fields can mutually benefit.
\end{abstract}

\begin{document}

  \maketitle

  \section*{Introduction}
  Rapid progress in \gls{ai} has resulted in both transformative advances in specialized domains and
  increasingly general-purpose foundation models and agents that can be effectively utilized
  across a large number of tasks. The applications range from protein
  folding~\citep{jumper2021highly}, weather prediction~\citep{lam2023learning}, algorithm
  discovery~\citep{novikov2025alphaevolve, yu2025autonomouscodeevolutionmeets}, and medical
  diagnostics~\citep{tu2024towards, nori2025sequential} to the more consumer-oriented vision of \gls{ai}
  personal assistants~\citep{gabriel2024ethics} that can perform diverse tasks by invoking the
  available tools. However, realizing the full potential of these systems requires addressing
  persistent challenges in reliability, explainability, and computational efficiency.

  Alongside \gls{ai}, major strides have recently been made in \gls{qc} both in terms of the
  underlying hardware, as well as theory and algorithms.  Platforms based on superconducting
  circuits~\citep{krantz2019quantum, castelvecchi2023ibm}, trapped
  ions~\citep{pino2020demonstration, monroe2021programmable}, neutral atoms~\citep{browaeys2020many,
  radnaev2025universal}, quantum dots~\citep{Harvey_2022}, and
  optics~\citep{psiquantum2025manufacturable} consistently demonstrate increased qubit counts,
  enhanced coherence times, and improved gate fidelities.  Experiments have now demonstrated the ability
  of quantum computers to exceed classical computers on some specialized tasks~\citep{arute2019quantum, morvan2024phase}; however,
  those tasks maintain some distance from useful applications.  Algorithmically, the major point of focus
  has been on developing methods to unlock applications in the \gls{nisq} era~\citep{preskill2018quantum}, and more recently also beyond, as systems transition towards
  fault-tolerant devices capable of \gls{qec} ~\citep{Preskill_2025}. 
  This includes the development of robust error
  mitigation~\citep{endo2018practical, cai2023quantum} and \gls{qec}
  techniques~\citep{knill2000theory, chiaverini2004realization, lidar2013quantum,
  bausch2024learning, google2025quantum} designed to transition from near-term
  to robust future devices.  Some of the promising near-term directions include quantum simulation~\citep{li2017efficient, barratt2021parallel,
  king2025beyond}, ground state approaches based on sampling for applications in quantum chemistry and materials science~\citep{robledo2025chemistry, kanno2026quantum}, and \glspl{qaoa}~\citep{farhi2014quantum, farhi2016quantum, zhou2020quantum} for combinatorial
  optimization problems. The simultaneous advancement of hardware capabilities and tailored algorithms is
  aimed at demonstrating significant strides in scientific and industrial problems.

  At their core, AI and quantum computing both seek to solve hard
  problems that are central to foundational questions in science. \gls{qc} aims to address these challenges by utilizing quantum phenomena that introduce
  new fundamental operations to develop more
  efficient algorithms with provable scaling advantages, whereas \gls{ml} aims to produce highly accurate
  approximations to otherwise computationally intractable quantities, e.g.~through the use of data. 
  The field of \gls{qml} aims to utilize the complementary strengths of both
  approaches~\citep{schuld2015introduction, biamonte2017quantum, havlivcek2019supervised,
  schuld2019quantum, cerezo2022challenges, jerbi2024shadows, bowles2024better,
  wang2024comprehensive}, seeking to synthesize these computational paradigms by leveraging quantum operations alongside both quantum and classical data.

  \gls{qml} develops learning methods centered around utilizing underlying quantum phenomena,
  such as superposition and entanglement. It also encompasses methods suited for directly manipulating quantum
  states from nature, making its scope far broader and more ambitious than simply
  adapting pre-existing classical \gls{ml} methods to run on quantum devices. For example, \gls{qml}
  can manipulate states or data from the physical world directly before performing a
  collapsing measurement, like the famous Schrödinger's cat, and when done this way, it is 
  said to be manipulating ``quantum data'' rather than the post-measurement ``classical data''.  
  In such cases, there are known examples of proven, unconditional exponential advantages 
  based on information theory that no amount of classical computation can overcome~\citep{aharonov2022quantum,chen2022exponential,huang2022quantum}.
  However, many of these advantages are on theoretical constructs, and establishing a
  definitive advantage on real-world problems has proven difficult. Quantum advantage
  is notoriously difficult to establish, and such claims can be subsequently invalidated by further advances in classical
  methods, referred to as "dequantization"~\citep{tang2019quantum, larose2024brief, huang2025vast}.
  Many of the early quantum supremacy claims~\citep{aaronson2016complexity, harrow2017quantum,
  boixo2018characterizing, zlokapa2023boundaries} involve constructions that may not translate to
  practical applications. For \gls{qml} to be truly useful, it needs to map onto meaningful
  problems with clear real-world utility, involving a shift from classically-inspired
  benchmarks onto scientific problems centered on intrinsically quantum data.  This shift would also
  help sidestep the substantial overhead incurred when loading large classical datasets into quantum
  states.

  Furthermore, we must be mindful of the potential trade-off between trainability and
  quantum advantage. Although recently the existence of beyond-classical, trainable generative quantum
  advantages have been proven and experimentally demonstrated~\citep{huang2025generative}, in some cases easily optimizable models
  may be more susceptible to efficient classical
  simulation~\citep{cerezo2025does}. Despite the challenges of loading classical data into
  quantum computers (often via constructions like Quantum Random Access Memory (QRAM)~\citep{giovannetti2008quantum} or Quantum Read-Only
  Memory (QROM)~\citep{babbush2018encoding}), some recent works are investigating how such constructions can be leveraged 
  with fundamental quantum mechanics to yield advantages in machine learning that survive, even if quantum computers
  turn out to be no faster than classical computers, also known as unconditional advantages~\citep{montanaro2024quantum,gilboa2024exponential,zhao2026exponential}.

  For \gls{qc} and \gls{ml}, the benefits are mutual~\citep{alexeev2025artificial}. 
  Indeed, it has been recently shown that adding data from quantum computers to classical training sets
  can provably increase the power of classical models beyond standard classical computation~\citep{huang2021power,huang2022provably,perez2024classical}.
  
  We begin by reviewing some of the existing contributions of classical \gls{ai} and \gls{ml} to the development
  of quantum technology.
  With that, we turn to \gls{qml} and discuss the challenges and opportunities for \gls{qml}
  to contribute to classical \gls{ml}, as well as newly emerging possibilities that are not available to purely classical systems.
  Finally, we conclude with an argument that
  closer collaboration between these two largely distinct fields would be highly beneficial, as it would establish the basis for a virtuous cycle to eventually unite them into a single quantum intelligence.

  \vfill
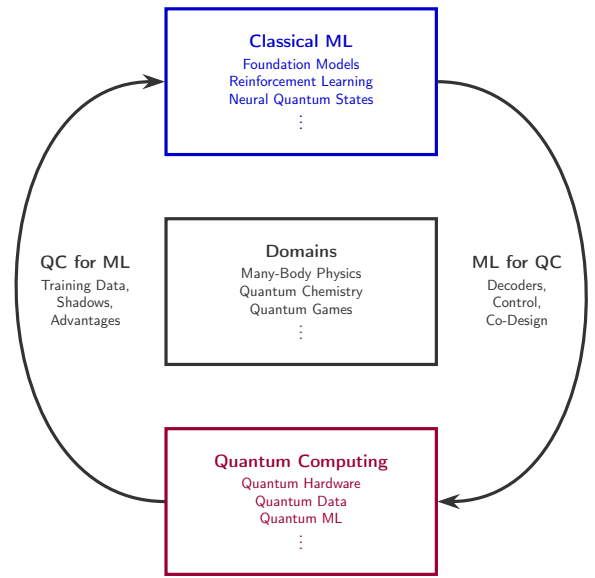
  \begin{figure}[h]
  \hspace{-3mm}\begin{tikzpicture}[
    scale=0.55,
    transform shape,
    node distance=1.5cm,
    box/.style={
        rectangle,
        very thick, 
        minimum width=6.5cm, 
        minimum height=3.5cm, 
        align=center, 
        font=\sffamily
    },
    cml/.style={
        box, 
        draw=blue!80!black, 
        text=blue!80!black
    },
    dom/.style={
        box, 
        draw=black!80, 
        text=black!80
    },
    qc/.style={
        box, 
        draw=purple!80!black, 
        text=purple!80!black
    },
    cyclearrow/.style={
        -{Stealth[length=3mm, width=2mm]}, 
        very thick, 
        draw=black!80
    }
]

\node[cml] (top) {
\textbf{\large Classical ML}\\[1mm]
Foundation Models\\
Reinforcement Learning\\
Neural Quantum States\\
$\vdots$};
\node[dom] (middle) [below=of top] {
\textbf{\large Domains}\\[1mm]
Many-Body Physics\\
Quantum Chemistry\\
Quantum Games\\
$\vdots$};
\node[qc] (bottom) [below=of middle] {
\textbf{\large Quantum Computing}\\[1mm]
Quantum Hardware\\
Quantum Data\\
Quantum ML\\
$\vdots$};

\draw[cyclearrow] (top.east) to[bend left=90, looseness=1.2] 
    node[left=0.4cm, align=center, font=\sffamily, text=black!80] {
        \textbf{\large ML for QC} \\[1mm] 
        Decoders,\\
        Control,\\ 
        Co-Design
    } (bottom.east);

\draw[cyclearrow] (bottom.west) to[bend left=90, looseness=1.2] 
    node[right=0.4cm, align=center, font=\sffamily, text=black!80] {
        \textbf{\large QC for ML} \\[1mm] 
        Training Data,\\ 
        Shadows,\\
        Advantages
    } (top.west);

\end{tikzpicture}
\end{figure}

\clearpage

  \setcounter{tocdepth}{\mytocdepth}
  \tableofcontents

  \section{ML for QC Systems}
    While quantum technology has advanced rapidly, in some respects it is still in its infancy with
    many challenging technological hurdles that are yet to be overcome. At the same time, classical \gls{ai} has seen an incredible
    boost in its ability to contribute to a multitude of applications across different fields. Here we review and discuss the cases where classical \gls{ai} has already contributed to the development of \gls{qc} and quantum technology, and where we expect it might make a positive impact in the future.

  \subsection[ML for QEC Decoders]{ML for Quantum Error Correction Decoders}

  Quantum error correction (\gls{qec}) and fault-tolerant quantum computation turn noisy physical qubits into fewer, higher-quality logical qubits by encoding information into larger Hilbert spaces~\citep{knill1997theory, lidar2013quantum, campbell2024series, putterman2025hardware}. There has been great progress recently in realizing error suppression in quantum devices~\citep{google2025quantum,radnaev2025universal}. \Gls{qec} approaches involve periodic measurements to identify error syndromes, which are then processed by decoders to determine the appropriate corrections and prevent logical errors. However, optimal decoding is, in general, NP-hard~\citep{berlekamp1978inherent, Hsieh2011NPhardness}, presenting two major challenges. First, code structure: promising families such as color codes have historically lacked efficient, high-accuracy algorithmic decoders. Second, device noise: quantum hardware manifests complex, correlated noise—including spatial and temporal correlations~\citep{zou2024spatially, seif2024suppressing}, leakage, and non-Markovian dynamics~\citep{biswas2025noise, jayashankar2023quantum}—which challenge standard decoding techniques~\citep{huang2019performance}. Furthermore, decoders must operate within strict real-time latency constraints to prevent error buildup. To address these challenges, a large variety of algorithmic decoders have been developed, broadly divided between general-purpose decoders that make few assumptions about the error model but are often computationally slow, such as exact or near-exact decoders~\citep{beni2025tesseract, cao2025exact}, and fast decoders tailored to specific code and error model structures~\citep{higgott2022pymatching, gidney2023chromobius, demarti2024decoding, berent2024decoding}.
  
  Utilization of \gls{ml} for decoding in \gls{qec} has progressed significantly, employing supervised learning~\citep{Torlai_2017}, reinforcement learning~\citep{andreasson2019quantum}, graph neural networks~\citep{hu2025efficient, lange2025data}, and other neural architectures~\citep{kim2020quantum, havstom2023machine, overwater2022neural, wang2023transformer}. The premise is that \gls{ml} decoders can learn strategies for new families of codes directly from syndrome data, without requiring a hand-tailored algorithm, while inherently learning the complex, hardware-specific physical noise characteristics of the device~\citep{zhao2023benchmarking}. AlphaQubit~\citep{bausch2024learning} demonstrated for the first time that \gls{ml} could decode the surface code to near-optimal accuracy up to intermediate code sizes—a task widely considered intractable due to the prohibitive training data requirements typically associated with neural approaches. Building upon this, AlphaQubit 2~\citep{senior2025scalable} extended these results to large code distances and historically difficult-to-decode families like color codes, while achieving sub-microsecond decoding throughput. Beyond planar topologies, neural approaches have also shown promise for scaling to complex architectures like quantum low-density parity-check (qLDPC) codes~\citep{Gu2026Scalable}. Beyond full decoding, data-driven models are also routinely applied to readout error mitigation, learning hardware-specific readout correlations to de-noise measurements~\citep{kim2022quantum, liao2024machine, baoqem}.

  Ultimately, the transition to advanced architectures—including high-rate, yoked, and qLDPC codes—introduces complex decoding landscapes with intricate topological features and high-degree check operators. Because these structures pose severe challenges for conventional algorithmic decoders, data-driven machine learning approaches represent a promising avenue for developing viable decoding solutions.

 \subsubsection{Speculative impact on quantum tech}
 The potential impact of \gls{ml}-based decoding in \gls{qec} is substantial, though many of its long-term benefits remain speculative. While the gap between heuristic and theoretically optimal decoders can be computed for small systems under idealized noise, assessing how much of this gap can be closed under realistic, low-latency constraints remains an empirical challenge. Moreover, the search for new codes with more logical qubits per physical qubit (higher rate), lower connectivity requirements, faster gates, better robustness to qubit dropout, etc. is often restricted to classes of codes that have known efficient, high quality, fast decoders. Thus, if \gls{ml} can expand the set of code families for which we have viable decoders, their impact will extend far beyond improving thresholds on known codes—it will fundamentally unlock the co-design and discovery of entirely new quantum codes and protocols, a topic we explore in the next section.

  \subsection[ML for Code Discovery]{ML For Code or Protocol Discovery}
 
 Beyond decoding existing quantum error-correcting codes, \gls{ml} is increasingly being applied to fundamental research on discovering new codes and protocols. However, quantum code discovery is notoriously difficult due to rigid algebraic structures, extreme combinatorial complexity, and the heavy reliance on expert domain knowledge. Recent advances in code design have leveraged \gls{rl}~\citep{olle2024simultaneous, su2025discovery}, agentic~\citep{cain2026shor}, and hybrid quantum-classical techniques~\citep{yanay2026learningbettererrorcorrection} to discover highly efficient error-correcting codes and protocols. In these approaches, optimization is formulated as a multi-objective search—maximizing distance and encoding rate while satisfying physical hardware constraints and minimizing the physical-to-logical qubit overhead required to achieve practical fault tolerance.

 \subsubsection{Speculative impact on quantum tech}
 Early work on practical quantum error correction focused heavily on geometries compatible with early fabrication and control capabilities, naturally prioritizing the surface code and 2D planar layouts. However, despite possessing highly favorable properties—such as efficient, high-quality decoders—these planar codes suffer from substantial physical-to-logical qubit overheads. Recent hardware developments, such as neutral-atom arrays, have unlocked flexible, long-range connectivity, renewing interest in a broader class of codes such as quantum LDPC (qLDPC) codes and accelerating progress toward large-scale applications like Shor's algorithm~\citep{cain2026shor, khattar2025verifiable, babbush2025grand}.
The design space for these codes is vast, and the performance constraints are strict. For example, finding a code with a high rate and large distance is only the first step. A viable code must also be co-designed with manufacturable and controllable hardware, admit logical gate implementations that do not introduce excessive latency, be decodable within real-time physical constraints, tolerate hardware component dropouts (atom loss, or imperfect qubit yield), and suppress logical error rates sufficiently to support computations requiring trillions of operations. If \gls{ml} can successfully navigate this high-dimensional design space, the practical benefits could be substantial, potentially reducing the timeline to logical quantum computing and delivering orders-of-magnitude speedups in logical runtime.

  \subsection{ML for Application Discovery}
  Despite the development of algorithmic primitives and basic quantum features that exhibit separation
  from classical computing, one of the greatest challenges has remained the identification of concrete,
  valuable applications where these (in many cases mathematically proven) speedups translate into a demonstrable practical advantage, 
  as articulated in detail in Ref.~\citep{babbush2025grand}.
  This often comes as a surprise to people observing the development of quantum computing from the outside,
  given the number of reported advantages in the media and existing algorithms.  For example, it's common to
  conclude that the existence of an exponential speedup in some linear algebra applications (like solutions
  of linear equations) would immediately imply advantages in domains like machine learning. Or, similarly, that quadratic speedups like Grover's search will yield immediate practical search advantages. As we discuss
  a bit later, when one examines the costs carefully for real problem settings, these practical advantages frequently
  diminish~\citep{aaronson2015read}.  Some notable exceptions where concrete applications have been worked out
  from infancy to near deployment include cryptanalysis, DQI for OPI, and quantum simulation~\citep{kivlichan2020improved,huggins2025fluid,khattar2025verifiable,low2025fast}.

  The reason for this challenge can be hard to concisely describe, as the precise reason advantages don't
  persist in real problems can differ from problem to problem. For example the need to load large amounts of
  classical data, or the condition number of matrices in large problems.  At a colloquial level, it might suffice
  to say many instantiations of problems in the real world do not appear sufficiently quantum shaped.  
  
  Perhaps more frustratingly, finding the hiccup on the path to applications is far from trivial.  It
  typically requires deep application domain expertise, as well as strong background in quantum algorithms.
  These collaborations take time to establish and bear fruit, and subtle insights or new quantum-inspired algorithms may quickly dash hope for showing a quantum advantage.  

  As quantum primitives become better understood, it is possible that the task of finding quantum shaped
  problems that fit for speedups end to end could be greatly aided by artificial intelligence with a comprehensive
  knowledge of many (if not all) fields of study documented today.  Such vast information retrieval and reasoning for establishing potential matches between algorithms and
  applications could be incredibly fruitful.  We note that this endeavor is quite distinct from the objective
  of finding quantum algorithms or new quantum circuits with the help of AI.  Instead, this direction 
  aims to find the square hole of quantum shaped applications to match the existing square pegs of
  known algorithms, primitives, or simple combinations therein.
  
  \subsubsection{Speculative impact on quantum tech}
  Despite the existential risk of so few concrete applications being precisely articulated, quantum
  computing continues to receive considerable support in both the public and private sectors.  However,
  ultimately the projected upside and return of quantum technology will be related to the value it can
  deliver either academically, commercially, or philosophically, and any new applications discovered
  here can bolster the support and development for years to come.  In addition, the earlier we discover
  potential new applications, the more time we have to optimize them and make them ready for deployment
  on early quantum computers, potentially helping shape the hardware technology along the way.
  
  \subsection{ML for Quantum Control}

  Quantum control optimization broadly refers to the task of finding analog controls for quantum systems that lie beneath
  the digital abstraction of the gate level, like time-dependent external fields (e.g.,
  laser pulses, microwave signals) for steering a quantum system towards a target
  state~\citep{shore2011manipulating}. The corresponding optimization landscape between control
  parameters and an outcome fidelity is fiendishly difficult; the mapping from control pulses to
  outcome fidelity is highly non-linear, resulting in a high-dimensional, non-convex space provably
  littered with "traps". Although theoretical works have suggested that with sufficient controls, the
  landscape becomes trap-free~\citep{pechen2011there,russell2016quantum} in analogy to the overparameterized regimes of deep networks, challenges are frequently experienced in practice.  Numerical optimization techniques such as gradient ascent pulse
  engineering~\citep{khaneja2005optimal, sorensen2018quantum} and Krotov
  methods~\citep{reich2012monotonically, morzhin2019krotov}, can be computationally expensive and
  lead to suboptimal solutions when modeling complex systems. Experimental noise also needs to be
  accounted for. 
  These challenges present a particular opportunity for \gls{ml} to improve quantum
  control.

  \Gls{ml} for quantum control has branched into two primary paradigms: black-box optimization~\citep{moon2020machine} and
  model-based learning. Initial studies demonstrated the feasibility of using \gls{rl}, including
  Q-learning and policy gradient methods~\citep{bukov2018reinforcement}, to design control pulses
  for tasks like state preparation and quantum gate synthesis. Genetic algorithms and evolutionary
  strategies have also shown promise for optimizing pulse
  shapes~\citep{zahedinejad2014evolutionary}. As neural networks present a natural choice for
  modeling quantum dynamics, deep \gls{rl} arises as a natural
  choice~\citep{niu2019universal, an2019deep, an2021quantum} for improving quantum control. Bayesian
  optimization has also been utilized for fine-tuning parametrized control
   pulses~\citep{lazin2023high}. \Glspl{pinn} can incorporate additional
   physical constraints to make the models more accurate~\citep{norambuena2024physics}.
   Reinforcement learning from demonstration has been shown to yield sample-efficient quantum control~\citep{li2025robust}, and can even stabilize physical control parameters in real-time by learning directly from quantum error detection events~\citep{sivak2025reinforcement}. Sample efficiency, noise robustness, and the ability to
  develop models that take into account the specifics of the hardware, remain important in the
  \gls{nisq} era.

 \subsubsection{Speculative impact on quantum tech}
  Analog controls that could benefit from \gls{ml} exist in essentially all quantum technologies at the moment, 
  but differ in their scope and ambition. For example, full system analog control is used
  in the schedule of quantum annealers or in the large scale analog simulation of quantum systems~\citep{daley2022practical}.  
  Advances in 
  these areas could potentially lead to earlier empirical demonstrations of quantum advantage for physical system 
  simulation or optimization due to the lower abstraction overhead with respect to digital systems.  Within digital
  systems aimed at error correction, it is typically a design principle that scalability can only be 
  achieved through error correction of modular
  components that are digitized / quantized (in a generalized quantum sense) in some manner, and hence analog controls
  are typically reserved to the component level to maximize scalability.  This includes improved implementations of 
  gates and measurements with lower error rates and faster operation time.  While these may seem restricted in scope,
  those improvements are magnified immensely by the error correction process, and the returns 
  can be exponential in the code distance as the 
  error rates drop farther below threshold~\citep{devitt2013quantum,sivak2025reinforcement}.  
  Moreover, one can expand the definition of component beyond gate to entire syndrome measurements where 
  even more flexibility might be found.
    \subsection[ML for Hardware Co-Design]{ML for Quantum Hardware Co-Design}

    \Gls{qas}~\citep{martyniuk2024quantum} has been predominantly framed as a software optimization problem, more specifically as the search for the optimal parameterized quantum circuit structure for \glspl{vqa}. As the field transitions toward early fault tolerance, hardware-software co-design~\citep{patel2024curriculumreinforcementlearningquantum, quetschlich2025mqt, liu2025hardware} presents a new opportunity for making advances by exploring the combinatorial possibilities within the corresponding hardware architecture, in relation to the desired circuit. This aims to address one of the primary challenges in scaling quantum computers: determining the optimal connectivity graph between qubits and pairing it with highly efficient error-correcting codes. For example, high-rate \gls{qldpc} codes~\citep{bravyi2024high} require orders of magnitude fewer physical qubits than standard 2D surface codes, but this comes at a cost of complex, non-local qubit connectivity that can be extremely challenging to implement without introducing severe crosstalk~\citep{zhang2025qplacer}, routing bottlenecks, and noise.

    By jointly optimizing the physical hardware layout and the error-correction codes, \gls{ai} agents can systematically discover favourable physical qubit placements, dynamic routing protocols, and modular inter-chip connections~\citep{tang2024alpharouter, zhou2026reinforcement}. These agents can empirically uncover the complex correspondence between the requirements of \gls{qldpc} codes and hardware topologies that are more easily manufactured~\citep{he2025discovering}. As these approaches mature, this type of co-design may prove to be one of the primary approaches for efficiently deriving optimal placements of non-local connections, and improving the overall system performance.
    
  \subsection[ML for Circuit Ansatz Search]{ML for Quantum Circuit Ansatz Search}
  Within \glspl{vqa} or parameterized quantum circuits,  
  the specific parameterized family of circuits being used is sometimes called the ansatz or architecture.  \Gls{qas}~\citep{du2022quantum, zhu2022brief} aims to automate the design of
  quantum circuits, to solve specific computational tasks efficiently, with the obvious short-term
  emphasis on \gls{nisq} devices due to present-day implementability. 
  Manual design of \glspl{vqc} may prove challenging due to the size of the
  search space and the number of possible choices in gate sets, qubit connectivity, circuit depth,
  and parametrizations.  Making good choices today requires deep intuition about the hardware and the
  nature of the problems being solved. In the future, when such methods are used on fault-tolerant devices,
  performance considerations as well as problem fit are expected to remain essential.
  \Gls{ml} methods may therefore be used to help achieve a more
  optimal circuit design, and help discover novel, high-performing, and hardware-efficient quantum
  circuits. Such advances would help accelerate research across quantum
  applications~\citep{cerezo2021variational}.

  \Gls{rl} methods may be used to iteratively construct and refine quantum circuits,
  in environments and tasks where a clear reward signal can be provided. These rewards may simply
  correspond to the accuracy of the associated task being run on the circuit. \Gls{rl} models have been
  used to select gates and improve their placement~\citep{kuo2021quantum, ye2021quantum,
  dai2024quantum}.  As in many other domains, evolutionary optimization offers an alternative where
  a population of promising solutions is maintained and evolved towards identifying optimal
  configurations. Individual mutations may involve adding or removing gates, and bits and pieces of
  different solutions may be recombined~\citep{ding2022evolutionary, zhang2023evolutionary}.
  Gradient-based methods have also been developed, as it may be possible to relax the discrete
  choices into a continuous space, allowing the use of
  backpropagation~\citep{zhang2022differentiable}. In many applications, adapting the solution to
  the specifics of the hardware and the associated noise levels is an important
  factor~\citep{cincio2018learning, murali2019noise, wang2022quantumnas}.

  The strengths of these traditional methods may be complemented by recent advances in \glspl{llm}, opening a promising new frontier in \gls{qas} and circuit synthesis~\citep{liang2023unleashingpotentialllmsquantum, nakaji2025generativequantumeigensolvergqe}. These models can be used to bridge the gap between the high-level algorithmic ideas and low-level instructions, by effectively performing code transpilation. Such applications may still benefit from specialized, fine-tuned models~\citep{jern2025agent}, that have been trained on large quantities of circuit diagrams and quantum code, though it is likely that these capabilities will keep improving in general-purpose models in the future. \Gls{llm}-driven approaches may still benefit from having access to traditional optimization loops, and involve simulators and reward feedback. As these technologies improve, we can expect them to dramatically decrease the amount of human effort needed for these tasks~\citep{gujju2025llm, macarone2026quantum}.

  To fully scale up \gls{qas}, the central bottleneck remains the prohibitive cost of evaluating candidate circuits. In particular, it is important to develop methods that can deal with the high cost of evaluating candidate circuits either via simulation or directly on quantum hardware. While it may be possible to achieve additional speed-ups in simulators, it would be imperative to develop \gls{ml} methods that are not only sample-efficient, but also generalizable. That way, it may be possible to iterate on the solutions at a smaller scale in order to identify more general and reusable patterns that can later be reassembled into more complex solutions at higher scales. To bypass the prohibitive cost of fully training each candidate circuit, recent methods employ zero-cost proxies, as metrics that estimate trainability and expressivity at
  initialization~\citep{he2025adaptive}.

 \subsubsection{Speculative impact on quantum tech}
  Much like the space of classical \gls{ml} and neural networks, there are many knobs one can turn within
  a \gls{vqa}. This includes the encoding from classical data into the quantum circuit, the architecture or ansatz that defines the model, the optimization algorithm, the output like a bitstring or expected value, and potential non-linear decoding of the output into an eventual prediction.  This space is vast, and determines aspects ranging from the trainability of the entire system to its ability to efficiently express functions that are not known to be efficiently expressible classically, and hence attain quantum advantage.  To date this space has been mired with a number of challenges, but if \gls{ml} can help identify the most promising directions, this may prove to be an opportunity for early quantum computers and \gls{ml} to be tightly coupled to achieve empirical beyond-classical demonstrations using real experimental data in training.

  \subsection{Neural Quantum States}
  Neural networks can be used to represent many-body wavefunctions, with downstream applications in
  condensed matter physics, quantum chemistry, and materials science. Such approximations are
  necessary, given that the Hilbert space of a system of $N$ quantum particles grows exponentially
  with $N$. For a system of $N$ spin-1/2 particles, the state vector requires $2^N$ complex
  coefficients, which quickly becomes intractable for classical computers. \Glspl{nqs}~\citep{jia2019quantum} provide a powerful and flexible variational ansatz for modeling
  these wavefunctions. The \gls{nqs} approach stems from a variational
  principle~\citep{PhysRevB.100.245123}, whereby for a wavefunction $|\Psi_\theta\rangle$, the
  expectation value of the Hamiltonian $\hat{H}$ provides an upper bound to the true ground-state
  energy $E_0$: $E(\theta) = \frac{\langle \Psi_\theta | \hat{H} | \Psi_\theta \rangle}{\langle
  \Psi_\theta | \Psi_\theta \rangle} \ge E_0$.  Minimizing this energy functional by finding the
  optimal parameters $\theta^*$, results in the best possible approximation to the ground state
  within the variational family defined by the chosen ansatz. In \gls{nqs}, neural networks are used to
  predict the wavefunction coefficients~\citep{carleo2019netket, carrasquilla2021use}. While this
  workflow is based on variational Monte Carlo, standard gradient descent is often unstable;
  consequently, more advanced optimizers like stochastic reconfiguration may be used instead~\citep{Chen_2024}. Many neural network models can be
  used in these applications, from \glspl{rbm}~\citep{deng2017quantum}, to feed-forward models~\citep{bausch2020quantum}, and
  transformers~\citep{lange2024architectures}. As physical wavefunctions must obey the symmetries of
  the Hamiltonian, equivariant networks render the modeling task much
  easier~\citep{malyshev2023autoregressive, lange2024architectures}, given that the network then
  doesn't need to derive and learn these symmetries on its own. Extensions have also been developed
  to handle the antisymmetric constraints of fermionic systems~\citep{keeble2023machine}. However,
  despite their versatility, training \gls{nqs} for large, strongly-correlated systems remains a
  formidable computational challenge.

  By applying the time-dependent variational principle~\citep{kramer2008review}, \gls{nqs} can simulate
  the real-time evolution of a quantum state under a given Hamiltonian~\citep{carleo2017solving,
  van2025many}. By projecting the Schrödinger equation onto the tangent space of the variational
  manifold, it is possible to obtain coupled differential equations for the network parameters
  $\theta(t)$. This has enabled studies of quantum quenches~\citep{donatella2023dynamics} and
  thermalization dynamics~\citep{PhysRevB.83.094431} in closed systems. However, numerical stability
  and error accumulation remain significant hurdles for long-time simulations. \Gls{nqs} can also be used
  for finding excited states~\citep{choo2018symmetries}, though this problem is more challenging as
  these states are not minima of the energy functional. These extensions may involve penalty terms
  to enforce orthogonality to the ground state~\citep{wheeler2024ensemble}, or a variational
  minimization of the energy variance.

  \subsubsection{Speculative impact on quantum tech}
  The ultimate goal of classical methods for simulating many-body quantum systems is typically either to attain predictive accuracy in modeling or greater mechanistic understanding that eventually results in the ability to control or improve that system or a related one.  For most digital quantum systems today, in order to achieve scalability in simulation and modeling of components, we typically model single or few-body quantum systems like single qubits or quantum oscillators under the drive of external fields, potentially coupled to a quantum bath.  However, beneath some of those models can be complicated many-body effects for which better understanding could unlock new insights to ultimately make better components. As an example, the issue of \gls{tls} impurities in superconducting qubits has long been the subject of debate with regards to its mechanistic origin~\citep{muller2019towards,wang2025superconducting}. Improved and advanced modeling of the material systems that comprise the qubits could reveal insights that could lead to changes in the fabrication or control process to remove or mitigate these effects.  Similarly, more advanced modeling could improve the accuracy of quantum control methods, when many quantum systems are thought to be inextricably involved in the entire component package, ultimately leading to improvements in the logical error rate.

  \section{Quantum Machine Learning}
  In broad terms, \gls{qml} is a strict generalization of classical \gls{ml},
  in that nominally it has many of the same goals, and of course with enough quantum resources any classical \gls{ml} algorithm can be run on a quantum computer through reversible arithmetic~\citep{nielsen2010quantum}.  
  Identical to classical \gls{ml}, 
  given data in the abstract sense, \gls{qml} seeks to learn and generalize the behavior of a given system.  
  Quantum operations open new
  opportunities to directly process systems in their natural quantum state, sometimes called quantum 
  data, which is not possible with a purely classical system.  
  However, like many tasks in classical computing
  there is not always a case for a quantum advantage, and for such tasks it is likely the more economical
  hardware, e.g., classical, is to be preferred.  Here we discuss some prominent approaches and insights
  from the field of \gls{qml}, and what they hope to contribute to \gls{ml} as a field.

  \subsection[Learning from Quantum Data]{Direct learning from quantum data}
   It is, in some sense, the definition of a quantum computer that it allows the manipulation of quantum states
   under complex coherent operations without collapsing the wavefunction to a classical state, unless desired.
   In many parts of quantum computation, we imagine taking classical input, and generating that state ourselves,
   and in such circumstances we must wonder if a classical computer could have accomplished the same task through
   some trickery.  In this situation, we imagine the unknown quantum state being delivered to us as fundamental
   elements from nature, and in such a case the ability to manipulate quantum states before measuring them
   adds a fundamental capability that can be proven as unconditionally more powerful than classical computers
   without this ability~\citep{huang2022quantum,aharonov2022quantum,cotler2021revisiting}.

   This raises the natural question of where quantum data comes from and what exactly it looks like.  Today,
   perhaps the most widely used quantum technologies in applications are quantum sensors~\citep{degen2017quantum} and clocks~\citep{altaie2022time}, which are
   arguably quantum sensors for time.  Although the field of quantum sensing has developed ways to use entanglement, squeezing~\citep{lawrie2019quantum},
   and other effects inside quantum sensors for e.g., Heisenberg scaling, most deployed systems today are based on
   single quantum particles or thermal ensembles of single quantum particles.  Recent proposals have interestingly
   shown that unique and new advantages may be possible when coupling even these single particles to full quantum computers~\citep{allen2025quantum,khan2025quantum}; however, the most general quantum data can come in the form of many-body quantum systems.

   It is these many-body quantum states where the largest advantages in quantum learning have been found.  In particular,
   even for quantum states of many particles with no entanglement, one can exhibit exponential separation in
   the number of samples required to learn about an unknown quantum system, rendering facets of the universe that would have
   otherwise been completely invisible, totally obvious~\citep{huang2022quantum,aharonov2022quantum,oh2024entanglement}.  However for such quantum data to be available to a quantum computer,
   objects like quantum sensors would need to be engineered to high fidelity with many qubits containing a multitude of information, or it would need to come from some other quantum computer or system we did not completely control.  Moreover, the process
   of transferring that quantum data into our logical quantum computer, 
   a process called quantum transduction~\citep{lauk2020perspectives}, can be challenging
   and needs to be improved between different platforms.  With improvements in those pieces, we may unlock totally new capabilities.

   To date, the advantages shown empirically in this area have largely focused on very minimal quantum operations
   in order to speed the path to implementation and understanding.  Eventually, however, we imagine that
   complex fully quantum models could be run on such systems, with benefits both to our ability to learn about
   and predict the behavior of ever more complex quantum systems.

  \subsubsection{Speculative contributions to \gls{ml}}
   This direction offers distinct theoretical contributions to \gls{ml}, despite the ambiguity in practical implementations.  The ability to manipulate quantum data directly adds a fundamentally
   new tool to the toolbox for learning about our physical world that does not exist in any form in classical
   computation.  On the other hand, the application of this tool to problems in the world today remains unclear, 
   for cases where there are known exponential advantages. It is our hope, that as classical \gls{ml} was guided
   by empirical exploration, so will be \gls{qml}.  By equipping researchers and general users alike
   with these capabilities, and improving the quality of quantum data sources, we will eventually come to understand
   what new avenues of exploration this technology unlocks.
  
  \subsection[Training Data from QC]{Training data from quantum computers}
  The rapid advancement in capabilities of classical \gls{ml} systems can make it hard to
  understand exactly where quantum learning will eventually play a role.  There are still
  some problems where quantum computers are expected to be able to do computations that no classical
  computation can do; however, the presence of data can provably change this frontier~\citep{huang2021power}.  
  For example, a number of results have now shown that despite the believed exponential advantages for direct
  simulation of quantum systems, if even simple classical models are provided with some training data
  as a restricted form of advice, these problems can become easy~\citep{huang2022provably,lewis2024improved,rouze2024efficient,wanner2024predicting}. The addition of this valuable data not only divides learning algorithms equipped with training data from quantum computers from regular
  classical computation, but provides a hint as to the potential benefits of quantum computers even in their
  infancy.

  To practically realize these benefits, quantum data needs to be appropriately extracted and represented for efficient classical processing. Full quantum state tomography would be prohibitively expensive, as it would require an exponential number of measurements. On the other hand, it is possible to instead aim to provide \emph{classical shadows} derived by randomized measurement protocols, and these sample-efficient classical descriptions can be used to capture the key properties of the underlying system~\citep{huang2020predicting}.

  The utility of quantum data for \gls{ml} applications may also depend on how it is accessed. For example, recent theoretical results have established unconditional exponential separation between learning with and without quantum memory~\citep{chen2022exponential}, subsequently followed by a demonstration showing strict learning separation utilizing just two copies of a quantum state at a time and the essential adaptive nature of subsequent measurements~\citep{chen2024adaptivity}. If a quantum state is measured immediately and the \gls{ml} model provided only the resulting classical data derived from this immediate measurement, the associated sample complexity may still be unfavorable for certain classes of tasks~\citep{gyurik2023limitations}. Having access to quantum memory, on the other hand, enables algorithms to store states and successfully accomplish tasks with an exponential reduction in sample size~\citep{huang2022quantum}. Yet, fault-tolerant hardware that would be needed to reach these theoretical limits remains out of reach in the near-term. The immediate applications may therefore need to rely on the measure-first paradigm. In such applications, the quantum computer would act as a physical data oracle, offloading subsequent learning and inference onto scalable classical resources. This allows for utilization of hybrid approaches and the development of highly expressive classical surrogate models~\citep{schreiber2023classical}.
  
  While this hybrid approach remains incredibly promising, it is still believed that even with training data, some problems like the discrete log or factoring remain hard without a quantum computer~\citep{Liu_2021,huang2021power}.  
  Hence we expect one 
  fruitful avenue of \gls{qml} to be represented by the improvement of classical models through data collected on quantum computers or quantum enhanced experiments~\citep{huang2022quantum}.

  \subsubsection{Speculative contributions to \gls{ml}}
  Among the contributions of quantum computers to \gls{ai}, we suspect this may be the earliest to have a real world impact.
  \Gls{ai} systems today are already used to great effect in chemical and material modeling, and the adoption is only
  expected to grow~\citep{keith2021combining,meuwly2021machine}. Within chemical and material modeling, there are problems known to be difficult to model classically, and difficult to precisely measure in experiments. If a quantum computer can gather enough high quality data at these critical parts of design space, it may dramatically improve the \gls{ai}'s ability to traverse that space
  accurately. The ability to generate quantum data for these problems could therefore turn out to be a compounding strategic advantage. Eventually this repository of training data is expected
  to extend into other \gls{ai} application areas. It is possible we will see other forays into this space, like quantum tool use in agentic systems both at training and inference time.

  \subsection[Unconditional Quantum Advantages]{Unconditional Quantum Advantages}
  For many practitioners, the fact that \gls{qc} enables an exponential number of complex amplitudes to be stored inside quantum states with $n$ qubits, highlights an opportunity to effectively store large classical databases in small quantum systems. Those hopes are then quickly dashed by Holevo's theorem~\citep{holevo1973bounds} and related results that show how despite 
  storing $O(2^n)$ complex amplitudes to high precision in an $n$-qubit state, it would only be possible to subsequently retrieve $O(n)$ classical bits on measurement. However, that turns out not to be the end of the story.

  Indeed, it was later found that despite the inability to extract all the information from such a state,
  there were tasks that when distributed between two players, the players could solve them by sending
  exponentially fewer qubits than bits, also termed an exponential communication advantage 
  for classical tasks~\citep{brassard2003quantum,buhrman2009non}.
  These types of unconditional communication advantage have recently been explored in the context of
  classical \gls{ml} applications~\citep{gilboa2024exponential,montanaro2024quantum}. Like communication advantages classically,
  the players in the game can be made into the analog of a single player in the past and present to
  manifest streaming advantages~\citep{le2006exponential}, where the problem specification is much larger than a player's practical
  memory and the information may only pass a bounded number of times in a particular order.

  To date, several quantum streaming advantages have been shown for different computational problems~\citep{kallaugher2022quantum}.
  At a glance this setup feels applicable to \gls{ml}, where it's not uncommon for training
  data sets to exceed the size of practical memory.  In order to prove rigorous separation,
  streaming advantages often assume a fixed, sometimes adversarial ordering, and noiseless queries to
  the data, unlike the typical setting of training.

  Recently these limitations were lifted, and an unconditional quantum space and sample advantage was 
  proven for a very general class of classical data problems, that can be sampled randomly and
  with noise~\citep{zhao2026exponential}. It develops a new technique for ensuring noisy data access does not destroy coherence
  that may be applicable elsewhere.  The main limitations of this approach are that the advantage is proven
  for a slightly more restricted memory than training set size, namely feature embedding size should not easily
  fit in memory.  Moreover, as a space advantage it is bound by the need to input all of the relevant data, so the
  runtime must still scale with the input size.  Barring the development of much faster quantum data loading
  technology, this may remain a barrier to practical implementations even in the future.
  
  \subsubsection{Speculative contributions to \gls{ml}}
  Unconditional advantages are exciting because they reflect the fact that the advantage is based
  on a fundamental tenet of quantum mechanics.  As such, assuming quantum mechanics is correct, when
  the conditions are met, even an infinite amount of classical compute cannot surmount them.  These conditions are restrictive, but not unimaginable.  If we take the present-day setting of finite context window
  \glspl{llm}, we can ask what the benefit might be of taking that same system and sitting a module next to it
  that could record every piece of information that ever passes through it for all time with fewer than $300$ logical
  qubits.  The questions one could ask on that data, and how many times one could ask, would of course be restricted,
  but the possibilities remain enticing.  Additionally, while memory may be abundant on earth, distributed systems
  we imagine deploying into space may be much more limited.  If such systems departed earth with only a few entangled
  qubits aboard, tied to $n$ qubits back here on earth, then they could record in the same way vast amounts of data
  and be able to teleport that state back to earth using a mere $n$ bits of classical communication for us to process
  here.
  
  \subsection{Variational Quantum Algorithms}

  Inspired by the success of classical neural networks, \glspl{vqa}~\citep{mcclean2016theory,cerezo2021variational} utilize
  \glspl{pqc} with trainable parameters that may also be
  referred to as the \emph{ansatz}, for quantum state preparation. This is followed up by a
  measurement of the prepared state, aiming to derive some desired properties or statistics. These
  derived properties of prepared states can then be used to compute classical cost functions that
  reflect how well each state approximates the desired solution to the underlying problem. The
  observed approximation errors can be used to update the \gls{pqc}, via a standard classical optimizer,
  minimizing the loss through an iterative process.

  While the short coherence times and heavy reliance on classical compute, sometimes referred to as
  its hybrid nature, of \glspl{vqa} make them a structural candidate for \gls{nisq} devices, their
  practical application is often hampered by significant measurement costs, trainability challenges, 
  and QPU-CPU latency. It
  is possible to keep the quantum computations shallow, complemented by more involved classical
  computations.  This heavier emphasis on classical computations may persist in early fault-tolerant
  devices considering the likely clock speeds. However, the efficacy of this loop is fundamentally
  limited by the communication overhead, requiring a tight integration of classical feedback, to
  avoid having the communication overhead dominate the runtime. Foundational examples of this
  approach include the \gls{vqe} for finding ground state energies of
  systems~\citep{peruzzo2014variational, belaloui2024ground} and the \gls{qaoa} for combinatorial optimization problems~\citep{farhi2014quantum,
  zhu2022adaptive}.  In the context of \gls{qml}, \glspl{vqa} can be employed as \glspl{vqc}~\citep{blance2021quantum, sen2022variational, zhou2023multi} or \glspl{qnn} more broadly. This includes the preparation of inputs as quantum states via appropriate
  feature maps~\citep{havlivcek2019supervised, suzuki2020analysis}. Input states are processed by
  \glspl{pqc} and the measurements of the resulting states correspond to model predictions.

  Despite the theoretical promise of \glspl{vqc} and structural similarities to successful classical deep networks,
  their practical utility is fundamentally challenged by a
  tension between expressivity and trainability~\citep{sim2019expressibility} and their underlying
  natural function space is quite different from their classical counterparts. As a result, training \glspl{vqc} can be
  quite challenging due to the emergence of barren plateaus~\citep{mcclean2018barren,
  larocca2025barren}, where gradients vanish exponentially with the number of qubits, and the
  proliferation of local minima~\citep{bittel2021training, anschuetz2022quantum}. More research
  is needed to better understand the trainability and expressivity of existing
  methods~\citep{sim2019expressibility}, especially since early studies have suggested that \glspl{pqc} may
  be able to express certain classes of complex functions with exponentially fewer
  resources~\citep{Du_2020}, and exponentially large Hilbert spaces may offer more expressivity in
  kernel methods~\citep{schuld2021supervised}. VQCs equipped with quantum kernels may be able to
  solve problems that are PromiseBQP-complete, and consequently all Bounded-Error Quantum
  Polynomial-Time (BQP) decision problems. Yet, this high expressivity is often the primary cause of
  the training difficulties, suggesting an inherent trade-off. Prior to the ability to run larger
  quantum calculations empirically, for an advantage to be meaningful, it
  needs to be demonstrated on a problem that is neither efficiently solvable by stochastic classical
  methods nor by quantum methods without learning, with the problem itself being of practical
  interest.  When empirical calculations are more readily available, other practical forms
  of advantage may be sufficient~\citep{schuld2022quantum, kim2023evidence}

  A flurry of recent research has sought to overcome \gls{vqc} limitations through heuristic extensions,
  including the utilization of hybrid optimizers~\citep{blance2021quantum, joshi2021evaluating,
  sen2022variational, acampora2023training}, improving the information
  encoding~\citep{Adhikary_2020, chen2020hybridquantumclassicalclassifierbased, yano2021efficient,
  thumwanit2021trainablediscretefeatureembeddings, wang2022development}, improving the
  ansatz~\citep{huang2022variational}, error mitigation strategies~\citep{li2024ensemble},
  architecture improvements~\citep{schuld2020circuit}, incorporating \gls{qram}
  circuits~\citep{duan2024parallelized}, and improving circuit width and
  depth~\citep{zhou2023multi}.  New frameworks are also being developed to improve the
  shot-efficiency in evaluating observables~\citep{liang2024artificial}, and a YOMO (You Only
  Measure Once)~\citep{liu2025measureoncedesigningsingleshot} approach has recently been proposed,
  as a departure from the more established Pauli expectation-value paradigm. In \gls{vsql}~\citep{li2021vsql}, classical shadows of quantum data are utilized, as
  auxiliary information tied to certain physical observables. Variational shadow circuits may be
  used to extract classical data via convolutions, reducing the required number of quantum gates.

  The general VQA framework admits many concrete neural network architectures; we survey the most prominent families in the following section.

  \subsection{Quantum Neural Networks}

  The development of \glspl{qnn} is motivated by the search for models that
  utilize exponentially large Hilbert spaces to surpass the limitations of classical deep learning.
  While some of the original attempts involved quantum annealing~\citep{dixit2021training}, the
  emphasis has shifted towards gate-based quantum circuits.  The hope is that \glspl{qnn} may offer computational speedups~\citep{wei2022quantum}, increased model capacity~\citep{Abbas_2021}, or improvements in training dynamics.

  \Gls{qnn} implementations on \gls{nisq} devices usually involve hybrid quantum-classical architectures,
  resulting in a hybrid approach with classical optimization~\citep{liu2021hybrid}. Various deep
  learning architectures have been adapted to this hybrid model. \Glspl{qcnn}~\citep{cong2019quantum, hur2022quantum, gong2024quantum} mimic the structure of
  classical CNNs by using layers of parameterized unitary operations followed by pooling. \Glspl{qcnn} can
  incorporate symmetries and inductive biases that make them well suited for some practical tasks
  like quantum phase recognition~\citep{herrmann2022realizing, umeano2025can}. Quanvolutional layers
  have also recently been introduced~\citep{henderson2020quanvolutional}, as local data
  transformations via random quantum circuits, similarly to the transformations performed by random
  convolutional filters~\citep{romberg2009compressive}. Quantum recurrent neural
  networks~\citep{bausch2020recurrent, li2023quantum, siemaszko2023rapid, nikoloska2023time, li2024quantum} may yield
  improvements in sequence-to-sequence modeling.  Quantum transformers~\citep{cherrat2022quantum,
  khatri2024quixerl, zhang2025survey} implement quantum attention mechanisms~\citep{chen2025quantum},
  and quantum diffusion models~\citep{cacioppo2023quantum, kolle2024quantum} can also be utilized
  for generative modeling. \Glspl{qgnn}~\citep{verdon2019quantum,
  collis2023physics} enable the modeling of structured data. \Glspl{dqc}~\citep{kyriienko2021solving} are well-suited for solving non-linear differential equations, of relevance in numerous scientific applications.

  As with all \glspl{vqa}, the trainability of specific \gls{qnn} architectures is shaped by the
  expressivity--trainability trade-off discussed above~\citep{qi2023barren}.
  Architecture-specific studies have begun to characterize where individual model families fall on
  this spectrum. In \glspl{qcnn}, the variance of the gradients decays no faster than
  polynomially~\citep{pesah2021absence, yang2025quantum}, suggesting better trainability than
  generic deep circuits. Yet, \glspl{qcnn} are also effectively classically
  simulable~\citep{bermejo2024quantum}, presuming access to the Pauli shadows on the dataset,
  reinforcing the link between classical simulability and ease of
  optimization~\citep{cerezo2025does}. One way for \glspl{qnn} to overcome these limitations would
  be to incorporate inductive biases relevant to quantum problems, building upon insights from
  geometric \gls{qml}~\citep{das2024permutation, jahin2025lorentz}. Equivariant \glspl{qnn} can be
  designed to commute with the symmetry group of the problem, making predictions invariant to
  particle rotations or permutations, and improving overall model
  performance. These methods also need to be rendered more scalable; for example, \glspl{qgnn}
  require qubit counts that increase linearly with graph
  size~\citep{ceschini2024graphsqubitscriticalreview}. Extending and improving these approaches
  for future fault-tolerant quantum hardware remains an important open direction.

  \subsubsection{Speculative contributions to \gls{ml}}
  While they are often not presented this way in order to facilitate near-term implementation in \gls{nisq} devices,
  \glspl{vqa} and \glspl{qnn} are a super-set and generalization of classical
  neural networks in the following sense.  Any classical neural network using $n$ classical bits of storage space
  with $T$ operations, can be efficiently mapped into a reversible computation on $O(n \log(T))$ classical bits 
  with $O(T^{1 + \epsilon})$ operations~\citep{bennett1989time}, where $\epsilon > 0$ is an arbitrarily small constant. This reversible circuit can
  be easily implemented in terms of unitary operations on a quantum computer, which upon bitstring readout gives a result
  provably identical to the original classical neural network.  One is then free to add operations to this network that
  are essentially quantum, e.g., a quantum Fourier transform, or general gate with parameters, and re-train the network.
  The function space this new circuit expresses, if the new element can be trained to the identity, is hence at worst
  as good as the original circuit, and potentially much better.  This type of mapping and example is obviously too
  resource intensive until we reach into fault tolerance, but it provides a reasonable basis for the future trajectory
  of \glspl{vqa} as being at least as powerful and trainable as classical neural networks with
  the ability to add operations that no classical computer is expected to be able to efficiently perform.  
  Moreover, it highlights several interesting deviations from the current paradigm of variational algorithms 
  that could make for interesting research directions as we develop new quantum hardware.
  
  \subsubsection{Generative models}
  One of the first empirical pieces of evidence for beyond-classical computations on a quantum computer
  centered around random circuit sampling~\citep{neill2018blueprint,arute2019quantum,morvan2024phase}. 
  In this task, there is an efficient classical description of a
  classical probability distribution that allows a quantum computer to easily draw samples, but it is believed 
  to be intractable (exponential time scaling in the number of bits) for any classical computer to draw
  samples from that same distribution with high accuracy.  Such demonstrations immediately raise the question
  of how to use this power for practical applications, and associated with this interest is a body of work
  on generative models on quantum computers~\citep{perdomo2018opportunities,liu2018differentiable,benedetti2019generative,wang2025towards,coopmans2024sample}.  
  However, the 
  same phenomenon in the distributions
  often linked to their hardness to simulate classically, appears to have a counterpart in making these
  distributions hard to train. This left a gap in our understanding of when advantages in generative
  modeling were possible, as efficient learning would be a requirement beyond just writing down a description
  at random one can sample from. One recent approach that aims to address this is that of \glspl{qcbm}~\citep{benedetti2019parameterized}, which trains parameterized quantum circuits as generative models in order to handle complex target distributions. Some of the distributions generated this way may also be intractable for efficient classical sampling~\citep{coyle2020born}.

  Recently, this loop was finally closed by identifying a classically hard to sample but efficient to
  learn distribution for a class of generative models~\citep{huang2025generative}.  
  These models take classical input and output,
  and are easy to evaluate and train for quantum computers, but hard to sample from with only classical
  computers.  While a substantial gap to practical applications remains, this work constituted both
  an existence proof for distributions like these as well as rigorous tools for identifying when 
  these models can represent empirical distributions without conflating that question with the difficulty
  of training.  It is expected that much like in classical \gls{ml}, tools like this used with
  empirical exploration will ultimately be the way to answer the question of which complex patterns can be described by these simple rules.
  
  \subsubsection{Speculative contributions to \gls{ml}}
   One aim of \gls{ml} and science more broadly is to explain complex behavior
   with a set of simpler rules that increase understanding and control over such systems. The question
   of whether there are natural and interesting classical datasets that are explained by simple quantum models 
   remains an open one, but recent advances in tools may help us begin to answer these questions empirically.
   With more quantum resources we may also be able to finally move beyond sampling in bits to sampling natively
   in distributions closer to our current natural descriptions of data sets such as with approximations of real
   numbers.  Much like the arguments listed for \glspl{qnn} generally, it is also possible to use
   a quantum computer to exactly duplicate the results of classical generative model with some small overhead,
   and to add some quantum operations to make it a superset of the available classical models that at worst stays
   the same.  However, the empirical observation that sampling experiments were the first to achieve beyond-classical
   performance in a lab, lends support to the conjecture that models rooted in these fundamentals may be some of
   the first quantum models to achieve advantages on practical distributions.  Much like how these questions
   about the learnability of real world distributions classically required massive empirical exploration, 
   we do not expect to be able to reliably predict via pencil and paper the exact crossover.  However, current
   evidence supports the idea that such empirical exploration will be worthwhile.

  \subsection{Quantum Reinforcement Learning}

  \Gls{qrl}~\citep{dong2008quantum, meyer2022survey} aims to leverage
  quantum effects to navigate complex state and action spaces, and more easily overcome bottlenecks
  in state space exploration. Current research is divided between near-term heuristic approaches and
  long-term fault-tolerant algorithms. In line with the other hybrid approaches, \gls{qrl} methods may
  utilize \glspl{vqc}~\citep{lockwood2020reinforcement} as function approximators in value functions or
  policies. This can be a swap-in replacement in a number of established \gls{rl} approaches, like \glspl{dqn}~\citep{mnih2013playingatarideepreinforcement},
  actor-critic~\citep{kolle2024quantumadvantageactorcriticreinforcement}, or policy gradient
  methods~\citep{10.1007/BF00992696, wang2022policygradientmethodrobust}. Quantum walks may help
  improve exploration in \gls{qrl}~\citep{Dalla_Pozza_2022}, and provably improving exploration has been a
  topic of broader interest~\citep{zhong2024provablyefficientexplorationquantum,
  santos2025extendingquantumreinforcementlearning}. Speedups in optimal action selection or Monte
  Carlo sampling efficiency may also result from the applications of Grover's
  algorithm~\citep{wiedemann2023quantumpolicyiterationamplitude} and quantum amplitude
  estimation~\citep{wiedemann2023quantumpolicyiterationamplitude}. \Gls{qrl} agents may also utilize \gls{vqls}
  as subroutines~\citep{Cherrat_2023}, or be integrated with evolutionary
  frameworks~\citep{chen2022variational}.

  Variational \gls{qrl} approaches remain affected by barren plateaus~\citep{Arrasmith_2021}, and the
  realizable qubit counts, gate fidelities, and decoherence, currently limit their applications to
  toy problems. For classical environments, these methods are currently crippled by the massive and
  unaddressed overhead in encoding environment states into quantum states, and in extracting this
  information through measurements. Action quantization in high-dimensional spaces is also
  challenging~\citep{Wu_2025}.  Identifying the types of
  problems~\citep{ghosal2024quantumreinforcementlearningnonabelian} where \gls{qrl} is likely to
  outperform classical \gls{rl} is crucial, and these are likely to be quantum environments, and problems
  involving quantum system control. Other environments where quantum advantage could be demonstrated
  have also recently been suggested~\citep{jerbi2021parametrized}, based on the discrete logarithm
  problem (DLP). Meanwhile, the development of more comprehensive benchmarks, better comparisons
  with classical methods~\citep{Kruse_2025}, and deepening the overall theoretical understanding of
  QRL~\citep{Dunjko_2017}, is needed to inform future innovation.

  \subsubsection{Speculative contributions to ML}
  A version of QRL that addresses purely classical problems with simple quantum model replacements of 
  e.g. the value or policy models is likely subject to the same speculation as previous sections in this
  domain.  Namely, that it is possible some complex policies or values are explained best by simple quantum
  models, but this is a question that will demand empirical exploration rather than pencil and paper conclusions.
  As such, it is fruitful to explore speculations tied more uniquely to reinforcement learning. For example, if
  learning with a natively quantum state space, then what we know about learning upon quantum data is immediately relevant.  Namely that there are manipulations that can be performed directly on the quantum
  state space to learn aspects of the state that would be invisible without these measurements, and manifest exponential differences that are insurmountable with computation~\citep{huang2022quantum}.  
  This suggests exponential benefits
  in natively quantum RL schemes achievable with very simple protocols as in previous works on learning
  on quantum data, though elaboration of these schemes into the domain of RL specifically remains a direction
  for future work.  Another uniquely RL aspect worth discussing is how it represents an alternative to the normal
  bottleneck of data loading in quantum machine learning.  In particular, slow data loading has led many to believe quantum computers require problem specifications that are small in size, but high in complexity
  to leave room for advantage.  The association of RL with game spaces that may be simple to specify, but require long evolutions to fully model may help satisfy this requirement of high complexity with low input sizes.

  \subsection{Quantum Kernel Methods}

  \Glspl{qkm} comprise a set of methods aimed at incorporating quantum computations
  with the aim of extending and outperforming classical kernel methods like \glspl{svm}~\citep{kavitha2024quantum} as well as performing classifications directly on quantum data.  
  More specifically, quantum computations are used in estimating
  the kernel function $K(x, x') = \langle \phi(x), \phi(x') \rangle$, where $\phi(x)$ is a feature
  map from a classical input to a quantum Hilbert space. In \gls{qml}, this mapping is normally realized
  as a parametrized quantum circuit, and the inner product can be evaluated efficiently on quantum
  devices~\citep{havlivcek2019supervised, schuld2019quantum}. Interestingly, the use of the so-called 
  ``kernel trick'' has not yet found a natural analog in fidelity-based kernels, 
  and the kernel similarities are evaluated by
  first explicitly mapping into a feature space defined by a quantum computer $\phi(x)$ before performing
  the inner product or analogue on the quantum computer~\citep{shin2025new}.  
  Where \glspl{qkm} are of particular interest,
  is in problems involving classically intractably large Hilbert spaces. It is also known that the
  quadratic quantum kernel is as expressive as an infinitely deep \gls{pqc} using an amplitude based encoding
  without data-reuploading~\citep{schuld2021supervised}.  However, the theoretical
  "efficiency" of \glspl{qkm} is often offset by a statistical sampling cost that scales polynomially with
  the desired accuracy, representing a hurdle for real-world applications.  In order to alleviate
  some of the challenges with storing training data in quantum memory as well as the sharpness of
  the fidelity-based metric of the original quadratic quantum kernel, \gls{pqk}
  methods have also been introduced~\citep{huang2021power}, where the kernel similarity is computed offline on classical shadows.  Although formally less powerful than the fully quantum models, they are still sufficiently expressive for some canonical beyond-classical  learning tasks related to the discrete logarithm, have reduced measurement costs, and generally more amenable
  numerical behavior.

  Modern feature map design has evolved along several specialized
  fronts~\citep{vlasic2025geodesicsquantumfeaturemaps}. For example, instantaneous quantum
  polynomial-time circuits are believed to be hard to be sampled from classically, inspiring the design
  of corresponding feature maps~\citep{havlivcek2019supervised, bremner2025instantaneous}. Feature
  maps may also be derived based on the preparation of a ground state of a parameterized
  Hamiltonian~\citep{albrecht2023quantum, umeano2024groundstatebasedquantumfeature}. It may be
  possible to avoid the high cost of encoding dense feature representations by instead mapping
  inputs onto a fixed set of Pauli strings, and computing predictions as their expected
  values~\citep{tiblias2025efficientquantumclassifierbased}. Quantum kernels can also be
  automatically discovered~\citep{incudini2024automatic} and trained~\citep{xu2024quantum}, as well
  as utilized as constituent parts of other methods, like the quantum kernel self-attention
  mechanism\citep{zhao2024qksan}.

  High dimensionality may potentially limit the generalizability of the derived models. However,
  this may be modulated in practice by varying the kernel width~\citep{canatar2022bandwidth,
  shaydulin2022importance}. Kernel width modulation should still be approached with a degree of
  caution, as these methods~\citep{egginger2024hyperparameter, schnabel2025quantum} may result in
  kernels that can be well approximated classically, standing in the way of achieving quantum
  advantage~\citep{slattery2023numerical}. The curse of dimensionality has other consequences as
  well, as it may lead to an exponential concentration of the values that the quantum kernel takes
  over its inputs~\citep{thanasilp2024exponential}, sometimes referred to as a "severe theoretical
  crisis" by skeptics as a manifestation of the barren plateau phenomenon. This concentration may lead to trivial models, where predictions are largely
  independent of the input data, unless specific care is taken.

  \Glspl{qkm} face a theoretical crisis arising from the inherent trade-offs, even though the
  formalism is of broader interest in \gls{qml}~\citep{schuld2021supervised}.  \Gls{qkm} problem formulations may
  also offer advantages in trainability, due to the convergence guarantees associated with convex
  objectives~\citep{hofmann2008kernel}. Despite theoretical advantages like convex trainability, the
  practical utility of \glspl{qkm} remains largely unrealized due to the prohibitive measurement cost.
  Achieving the desired level of precision in kernel estimation may offset other theoretical
  speedups. \Glspl{qkm} have been explored in the context of regression, differential
  equations~\citep{paine2023quantum}, and predicting quantum phases of
  matter~\citep{sancho2022quantum, Wu_2023}.

  \subsubsection{Speculative contributions to ML}
  Despite the rise of deep learning classically, kernel methods still play a role in a range of learning
  problems including the small data regime, problems with deeply known mathematical structure, 
  cases where training is especially challenging, and perhaps interestingly in the optimization of deep learning
  hyperparameters.  By extension, we might expect for \gls{qml} that \glspl{qkm} may
  play a role in similar circumstances.  Like with other \gls{qml} methods on classical data, the speculation remains
  that there may exist quantum kernels that naturally represent complex data.  If such is the case, kernel methods
  may play a role when very little data is available or no trainable \gls{qnn} of similar quality can be found.  The
  ability to directly use quantum data at both training and inference time may lead to simpler constructions
  for powerful models on direct quantum data prediction as well.

  \subsection{Quantum Linear Algebra}

  Linear algebra features prominently in classical \gls{ml}, and
  \gls{qla}~\citep{lipton2014quantum, zhao2021compiling} is equally central to \gls{qml}
  theory. However, a persistent challenge shadows this entire category: performing fundamental
  linear-algebraic operations on classical data typically requires a data-loading step via
  constructions like \gls{qram}~\citep{giovannetti2008quantum} or
  \gls{qrom}~\citep{babbush2018encoding}, whose cost scales with the dataset size and often
  negates any potential exponential speedup. Extracting classical information from the quantum
  output is also generally hard~\citep{aaronson2015read}. While this problem casts a long shadow
  and often necessitates \gls{qec}, interest remains in how quantum computers might provide
  advantages if these barriers were to be overcome.

  The \gls{hhl} algorithm~\citep{babukhin2023harrow} was one of the first demonstrations of
  quantum advantage for linear systems, with subsequently improved
  algorithms~\citep{childs2017quantum} achieving runtimes polylogarithmic in dimension. More recently,
  there have been innovations in quantum linear systems including matrix equations that allow
  totally different access and readout cost models with different tradeoffs and potential
  for novel quantum advantages~\citep{somma2025,wang2026sign}.  However,
  even improved \glspl{qls} face severe caveats: polynomial dependence on condition number
  and precision~\citep{harrow2009quantum}, ill-conditioned matrices common in \gls{ml}
  tasks~\citep{saarinen1993ill}, and the data-loading overhead noted above. \Glspl{qls} have been
  utilized in quantum support vector machines~\citep{rebentrost2014quantum} and Gaussian process
  regression~\citep{zhao2019quantum}. Similarly, \gls{qsvd}~\citep{bellante2022quantum} and
  \gls{qpca}~\citep{li2021resonant, xin2021experimental, he2022low} were introduced with the hope
  of exponential speedups for dimensionality reduction~\citep{lloyd2014quantum}, but
  \emph{dequantization} results~\citep{tang2022dequantizing} have shown that classical methods can
  achieve comparable performance in certain problem classes~\citep{tang2019quantum}, and the
  original speedup claims may be artifacts of state preparation
  assumptions~\citep{tang2021quantum}. Notably, \gls{qpca} retains a provable exponential advantage
  that cannot be dequantized in the presence of quantum data~\citep{cotler2021revisiting}.

  As a \gls{nisq}-era alternative, \glspl{vqls}~\citep{bravo2019variational} use \glspl{pqc} to
  approximate solutions to $Ax=b$ via classical loss minimization, and have been applied to fluid
  dynamics~\citep{turati2024empiricalanalysiseffectivenessvariational,
  bosco2024demonstrationscalabilityaccuracyvariational, rao2024performance} and
  PDEs~\citep{sarma2024quantum}. Related variational eigenvalue
  approaches~\citep{peruzzo2014variational} can be noise-resilient through shallower circuits, and
  the ansatz can be dynamically expanded~\citep{patil2022variational}. However, \glspl{vqls} remain
  subject to barren plateaus~\citep{mcclean2018barren}, noise-induced optimizer
  degradation~\citep{pellow2021comparison}, and the general data-loading
  overheads~\citep{dunjko2018machine}. One strategy to circumvent data loading entirely is to find
  problems naturally defined by lifting a compact description into a much larger space: tensor PCA
  has yielded a quartic speedup this way~\citep{Hastings_2020}, and topological data
  analysis~\citep{lloyd2015quantum} exploits the combinatorial graph Laplacian derived from a
  simpler data graph, though recent analysis suggests the resulting speedup is typically restricted
  to small polynomial factors~\citep{berry2024}.

  \subsubsection{Speculative contributions to \gls{ml}}
  Despite initial promise and hope, to date no end-to-end exponential speedup is known for a specific
  problem using quantum linear algebra routines.  The reasons for this are often nuanced, like the natural
  scaling of the condition number or readout, or more typical like the cost of properly accounting for data
  loading inside the quantum oracle.  However, asymptotics are not really the end of the story.  If we
  can imagine the development of a truly fast data loader such that coherent queries to a large database
  are negligible in wall clock time, then it is conceivable that one could find more widespread application
  of these techniques for real problems.  It is also certainly true that these methods have found use within
  other quantum algorithms, where the problems of data loading are ameliorated significantly.  If these challenges
  could be overcome, then we might expect more widespread speedups than are easily anticipated today.

  \subsection{Quantum Optimization}
  A topic of great interest in both classical and quantum \gls{ml} is training algorithms
  and finding the best models for a particular problem.  It has long been speculated that quantum
  computers may be better at hard optimization problems than their classical counterparts, and, indeed,
  this was the motivation for some groups to look into the use of \gls{qa} or more broadly quantum
  adiabatic computing as they might be applied even to classical \gls{ml} models to improve
  either training time or generalization quality.  The methods proposed for using quantum computers have
  varied over time, but given the broad applicability to problems across many domains including \gls{ml}, the overall interest has never faded.

  \Gls{qa}~\citep{finnila1994quantum, morita2008mathematical} is a heuristic
  optimization technique inspired by the formal framework of adiabatic quantum
  computation~\citep{farhi2000quantum, farhi2001quantum}, utilizing quantum tunneling to navigate
  towards lower energy states. It is specifically aimed at problems that can be mapped to finding
  the ground state of an Ising model~\citep{kadowaki1998quantum}, or problems that qualify as
  \gls{qubo}~\citep{punnen2022quadratic,
  PhysRevApplied.18.034016}. The hope is that \gls{qa} would be able to outperform its classical
  counterpart, simulated annealing, and more easily identify the desired target states. \Gls{qa} starts by initializing a set of qubits as a ground state of a transverse field
  Hamiltonian. This system is then evolved by gradually turning down the transverse field while
  simultaneously turning up the problem Hamiltonian, whose ground state encodes the solution to the
  optimization problem~\citep{kadowaki1998quantum}. \Gls{qa} can be useful in those \gls{qml} methods that allow
  for a \gls{qubo} formulation, which includes methods like Boltzmann Machines.

  Quantum annealers are often presented as more noise-resilient and scalable than their gate-model
  counterparts, as they do not require long coherence times or complex error
  correction~\citep{yulianti2022implementation}. However the lack of coherence time is often 
  seen by critics as evidence that the same processes could be emulated more cheaply with
  entirely classical devices.  That said, lower requirements permitting larger devices makes 
  \gls{qa} an appealing approach on near-term
  quantum hardware. \Gls{qa} has been applied to training Helmholz Machines~\citep{benedetti2018quantum},
  \glspl{rbm}~\citep{dixit2021training}, performing feature
  selection~\citep{nath2021quantum, pomeroy2025quantum}, and clustering~\citep{kumar2018quantum,
  neukart2018quantum, zaiou2021balanced}. \Gls{qa} devices can also support sampling from Gibbs
  distributions, which can be utilized by generative \gls{ml} models~\citep{amin2018quantum}.

  The primary hurdle for \gls{qa} remains the persistent competitiveness of classical
  heuristics~\citep{katzgraber2014glassy, mandra2017pitfalls, quinton2025quantum}. Limited qubit
  connectivity on current devices may get in the way of realizing speedups. Decoherence and noise
  may lead to thermal excitations away from the intended ground state. Most importantly, classical
  heuristics remain highly competitive~\citep{katzgraber2014glassy, mandra2017pitfalls,
  quinton2025quantum}. This includes tensor network methods like matrix product states, that are
  often capable of solving the optimization problems more efficiently than current quantum
  hardware~\citep{luchnikov2024large}.  Reverse Annealing aims to address some of the identified
  bottlenecks by starting from a classical state and exploring the landscape
  locally~\citep{venturelli2019reverse, Jattana_2024, PhysRevA.101.022331}. Despite the limitations,
  \gls{qa} remains useful as a specialized optimization and sampling method in the \gls{nisq} era.

  When formulated on a gate model quantum computer, it is common to change formulations from
  \gls{qa} to quantum adiabatic optimization~\citep{Reichardt2004,farhi2001quantum}, 
  where the formulations are similar but
  some of the crucial details may be changed. For example, the continuous time evolution of the Hamiltonian
  defining the approach is discretized and implemented by some fault-tolerant compatible approach, like
  Trotter factorization or otherwise~\citep{Sanders2020}.  
  Similarly, the driving or problem Hamiltonians can be made more general
  than Hamiltonians natively realizable on hardware, and the focus is often on finding a schedule or time that 
  provides rigorous performance guarantees.  This flexibility allows one to address more general optimization
  problems than are easily encodable in standard \gls{qa} setups, and hence may be more generally
  applicable to problems like training models in \gls{ml}.

  If one takes the Trotter digitized formulation of either \gls{qa} or quantum adiabatic optimization,
  and lets the parameters define a simple trainable path, then the result is the quantum alternating operator
  ansatz (\gls{qaoa})~\citep{farhi2014quantum}.  Despite similarities in the formulations and implementations, there are important differences.  For 
  example, problems with substantial separation between quantum and classical performance have been found
  that utilize very non-adiabatic mechanisms when examined carefully~\citep{bapat2019bangbang,McClean_2021}.  
  For near-term and early fault tolerant
  implementations the cost is very easy to control.  Also, for moderate depths, the \gls{qaoa}, the algorithm has the
  interesting and powerful property that performance can be efficiently predicted on a classical computer, but
  the solutions cannot be drawn efficiently without a quantum computer.  The flexibility and amenability
  of the method to study may allow it to lend some of the earliest benefits to \gls{ml}; however,
  it, like many quantum optimization algorithms, is typically formulated in terms of binary problems,
  and one may need to translate to more qudit-based representations to naturally address problems in \gls{ml}.

  As the landscape has developed further, a recent method offering mechanistically very different advantage opportunities for optimization has arisen called \gls{dqi}~\citep{Jordan_2025}. \Gls{dqi} offers speedup for approximate problems, and a \Gls{dqi} speedup has been shown on the problem of optimal polynomial interpolation, which is deeply related to \gls{ml}. The exact kind of problem structure that would be amenable to speedups with \Gls{dqi} is presently unknown, and remains an open problem. These recent developments offer renewed hope that quantum optimization could play a role in the training of \gls{ml} models, but much work is left to be done.

  \subsubsection{Speculative contributions to \gls{ml}}
  The wide applicability of optimization problems is a double-edged sword in their study.  On the one hand,
  meaningful advantage has pervasive impact. On the other hand, because the problems are so general, widespread 
  exponential advantage would violate widely believed theoretical conjectures like P$\neq$NP, or other
  notions like the no free lunch theorem.  Hence very large advantages must depend on some kind of structure, but
  those structures are still, as of yet, largely unknown.  However, if we start to develop this knowledge, assisted
  by empirical exploration, evidence suggests there may be problems where quantum computers offer advantages in time to solution, or quality of solution.  While training error, or its equivalent, has largely been
  the focus of quantum advantages, the properties of quantum diffusion processes lead some to conjecture there
  may be inherent advantages in generalization errors as well, or at least a better diversity of solutions.  Ultimately,
  this will be determined empirically, but any advantage on training could have immense benefits to \gls{ml}.

  \section[Synthesis]{Synthesis: Where Directions Meet}
  The preceding sections surveyed how classical \gls{ml} accelerates quantum technology
  development and how quantum resources create new possibilities for
  learning. These two directions are not independent: in several application domains they
  form a closed loop, where progress on one side directly enables progress on the other. We
  highlight three such domains---many-body physics, quantum chemistry, and strategic games---as
  case studies of this mutual reinforcement, and use them to identify cross-cutting patterns that
  span the methods discussed above.

  \subsection[Many-Body Physics]{Quantum Simulation of Many-Body Physics}
  Quantum simulation was perhaps the first proposed application of quantum computers, where a quantum computer
  plays the role of a more controllable version of a quantum system one wants to study.  Within quantum simulation
  there are many distinct domains and problems one can articulate with different methods and challenges.  Here 
  we talk about a few in the context of how they live in the \gls{qml} ecosystem.
  This domain draws on neural quantum states for classical wavefunction modeling and
  on quantum-generated training data for enhancing classical surrogates, illustrating
  both directions of the \gls{ml}--\gls{qc} interaction.

  Quantum many-body physics aims to model the collective behavior of systems composed of a large
  number of interacting quantum particles, such as electrons in materials, ultra-cold atoms in
  optical lattices, or spins in magnetic systems.  Quantum physics, as opposed to chemistry as discussed
  below, is often interested in the emergent behavior of simple systems into complex but robust 
  phenomena.  A central challenge in this field, especially with regards to simulation, is the
  ``exponential wall''~\citep{fulde2017dealing} arising due to the exponential increase in the
  dimension of Hilbert space based on the number of particles being modeled.  Exact solutions 
  on a classical computer are therefore only tractable for very small systems, in general. 
  This limitation in terms of scale stands in the
  way of a deeper and more comprehensive understanding of important physical phenomena like
  superconductivity, quantum magnetism~\citep{schollwock2008quantum}, and fractional quantum Hall
  effect~\citep{jain1990theory, stormer1999fractional}.  Established classical numerical approaches are known
  to face numerous obstacles in these domains. For example, quantum Monte Carlo methods run into
  the sign problem when dealing with fermions~\citep{troyer2005computational, li2015solving, li2019sign,
  zhang2022fermion}, whereas density matrix renormalization group~\citep{schollwock2005density,
  verstraete2023density} faces issues scaling beyond one-dimensional
  problems~\citep{schollwock2011density, PhysRevLett.133.190402}. There is therefore a pressing need
  for novel computational paradigms to address these obstacles, and \gls{qml} may prove to be an important tool in modeling these important systems, especially when equipped with efficient simulation methods provided by quantum computers.

  The earlier mentioned \glspl{nqs}~\citep{jia2019quantum} present a promising
  approach for expressing the many-body wave function amplitudes~\citep{carleo2017solving,
  medvidovic2024neural}. In these variational approaches~\citep{PhysRevB.100.245123}, the aim is to
  minimize the expectation value of the Hamiltonian, thereby approximating the ground state energy
  and wave function. The expressivity of neural networks and their ability to identify subtle
  patterns makes them well-suited for modeling the arising entanglement~\citep{deng2017quantum}.
  Beyond ground states, \gls{qml} can also be applied to model phases of matter, and, consequently, phase transitions.  However, the bar for demonstrating a distinct quantum
  learning advantage remains high; more work is needed to address optimization challenges and scale to larger Hamiltonians.

  These problems have also been approached using classical \gls{ml}~\citep{huang2022provably}, potentially empowered by quantum data.  It is worth
  distinguishing between the study of dynamics, ground states, excited states, and thermal
  states of matter.  Each of these has their own challenges and requires specific methods;  notably,
  quantum computers are provably efficient at simulating dynamics.  This efficiency may provide unique
  opportunities to augment classical \gls{ml} methods with data from quantum simulations, improving their accuracy.  

  \subsection[Quantum Chemistry]{Quantum Simulation in Quantum Chemistry}

  Quantum chemistry is a particular subset of the more general many-body physics problem.  Quantum chemistry focuses specifically on electronic problems, often in the field of classical nuclei, and with a particular focus on quantitative accuracy in the energies of particular systems
  rather than the emergence of general qualitative behavior.  This makes some of the approaches and goals
  quite different.  Specifically, the central challenge of quantum chemistry is solving the electronic Schrödinger equation to
  predict molecular and material properties~\citep{levine2009quantum}.
  Quantum chemistry sits at the intersection of variational quantum
  algorithms, \gls{ml}-derived surrogate models, and the paradigm of
  quantum-data-augmented classical learning.
  An accurate understanding of
  electronic structure plays an important role in understanding chemical bonding, reaction
  mechanisms, and other material properties. Computing the exact solutions to these problems is
  intractable for larger systems, and this presents an obstacle to furthering our understanding of
  materials. Classical high-accuracy methods like configuration interaction and coupled
  cluster~\citep{crawford2007introduction} have fairly steep polynomial scaling ($N^6$ or worse),
  and remain applicable only to small molecules.  \Gls{dft}~\citep{orio2009density} presents a more computationally feasible approach, scaling as $N^3$
  to $N^4$, making it a method of choice for slightly larger systems. Yet, its accuracy depends on
  having good approximations to the \gls{xc}
  functional~\citep{scuseria2005progress, peverati2012exchange, fishman2013accuracy}, and improving
  the quality of this approximation is an open challenge.

  Classical \gls{ml} is being applied at multiple levels of abstraction in quantum chemistry to bridge
  the gap between accuracy and scalability. Instead of directly solving the complex many-electron
  Schrödinger equation, \gls{ml} techniques can be applied to learn the underlying relationships, by
  generalizing from the data produced by more accurate and more expensive quantum chemistry
  simulations on smaller and more tractable systems.  For example, \gls{ml}-derived \glspl{pes}~\citep{jiang2020high, dral2020hierarchical, kang2020large, mills2022exploring} can
  be used to derive the relevant quantities for previously unseen configurations at a fraction of
  the usual cost. This, in turn, makes large-scale molecular dynamics simulations possible.  A key
  component in current methods is the encoding of the local environment in ways that respect the
  underlying physical symmetries. \Gls{dft} can also be improved by utilizing \gls{ml}, and neural
  networks more specifically to learn the \gls{xc} functional~\citep{dick2020machine,
  kanungo2025learning,Kirkpatrick2021}.  Direct representation of the many-electron wave function using neural
  networks, similar to the \gls{nqs} approach in many-body physics, is also being actively pursued in
  quantum chemistry~\citep{pfau2020ab}.

  In the \gls{nisq} era, the \gls{vqe}~\citep{peruzzo2014variational} has
  emerged as one of the most prominent \gls{qml} approaches for electronic structure calculations, along with sample-based quantum diagonalization~\citep{robledo2025chemistry, piccinelli2026quantumchemistryprovableconvergence}. \Gls{vqe} prepares a
  trial wavefunction for the molecule, and the expectation value of the molecular electronic
  Hamiltonian with respect to this trial state is subsequently estimated through measurements. The
  optimization process aims to identify the corresponding ground state. The effectiveness of \gls{vqe} in
  quantum chemistry depends on the choice of the ansatz. Chemically-inspired ansätze, such as the
  unitary coupled cluster singles and doubles~\citep{romero2018strategiesquantumcomputingmolecular}
  build upon successful classical quantum chemistry methods, though achieving deep circuits may
  still be challenging on \gls{nisq} devices. Hardware-efficient ansätze may sidestep this challenge, but
  otherwise be harder to optimize in absence of constraints encoding the chemical intuition. Other
  strategies like ADAPT-\gls{vqe}~\citep{grimsley2019adaptive} aim to grow the ansatz dynamically. Active
  learning~\citep{c8q9-6vy7}, may also be used to
  adaptively select the points at which to run \gls{vqe} on the energy surface, reducing the overall
  quantum runtime by a potentially large factor.

  Longer term, we imagine that data from highly accurate quantum computation can be used to
  further improve classical ML models both for potential energy surfaces, and even reduced
  models of electronic behavior.  Conversely, the improvement of these models may improve the
  modeling of quantum computers in a way that improves their quality and qubit count, leading to
  ever larger and more accurate quantum simulations to continue driving the classical ML models.

  Eventually, we imagine for truly quantum systems, native quantum models will be found and implemented fault-tolerantly. Due to the lower abstraction overhead
  they are likely to be more compact and efficient than their classical counterparts, especially 
  when quantum dynamics are involved. Of course, such conjectures will have to be tested empirically.

  \subsection[Quantum Games]{Games with Quantum Resources}
  Games have historically served as critical benchmarks for the development of new \gls{ml}
  techniques, and curiously were among the first domains in which unconditional quantum advantages were
  shown.  The conditions for players with quantum resources to show unambiguous and large advantages over their
  competitors with classical resources are restrictive, but not impossible to imagine~\citep{ding2025}.  For example,
  in most cases, the players must be unable to communicate for some reason like latency, they must make 
  local observations unknown to the other player, and be working to optimize some joint objective that is 
  non-separable. This has not prevented conjectures of how this might be applied in domains like finance, but
  the lack of very obvious applications has led some to wonder what we are missing.  Empirical playing of
  games with quantum resources may help shed light on this.
  Quantum games connect quantum reinforcement learning, unconditional communication
  advantages, and the broader question of where shared quantum resources yield strategic
  edges unavailable to purely classical agents.

  Broadly, games enable us to more easily isolate and investigate different dimensions of difficulty, and focus on developing specific methods and capabilities to overcome them. Popular
  games like chess~\citep{silver2017masteringchessshogiselfplay,
  schut2023bridginghumanaiknowledgegap, zahavy2024diversifyingaicreativechess},
  Go~\citep{silver2016mastering}, and poker~\citep{moravvcik2017deepstack, brown2018superhuman,
  brown2019superhuman}, have been an inspiration for numerous methodological advances over the years.
  Unsurprisingly, there is a growing interest in quantum games~\citep{benjamin2001multiplayer,
  li2002continuous, khan2018quantum} and quantum game theory~\citep{eisert1999quantum,
  flitney2002introduction, gutoski2007toward} as rigorous benchmarking environments. While it remains a speculative question whether these
  environments will yield direct breakthroughs for \gls{qml}, they provide a valuable framework for investigating quantum advantage in strategic decision-making and for evaluating \gls{qrl} agents. By enriching classical rule sets with quantum mechanics and quantum resources, these environments force learning agents to navigate state spaces characterized by superposition, entanglement, and general shared quantum resources.

  Several classical games have been quantised and introduced as QRL environments. This includes quantum tic tac toe~\citep{goff2006quantum}, quantum chess~\citep{cantwell2019quantum}, quantum checkers~\citep{11114278} and quantum poker~\citep{fuchs2020quantum}. These adaptations introduce new, uniquely quantum elements, such as superposition or entanglement, to their rule sets and action spaces. Quantum prisoner's dilemma has also been studied~\citep{debrota2024quantum, makram2024time}, and some Bayesian games have already been demonstrated on early quantum hardware~\citep{Solmeyer_2018}. \Glspl{pqc} were also suggested as a way of training AlphaZero on quantum devices~\citep{10.1145/3583133.3596302}. Beyond their primary utility as \gls{rl} benchmarks, many of these games concurrently serve an educational purpose~\citep{goff2006quantum, fuchs2020quantum, weingartner2023quantum} by introducing the broader community to the principles of \gls{qc}.

  The study of quantum game environments through quantum game theory has practical implications that extend beyond algorithm benchmarking~\citep{sanz2025mapping}. Insights derived from optimal multi-agent coordination in quantum games may help inform the optimization of decentralized quantum networks~\citep{10628033}. There are also possible applications in challenging classical problems, related to optimal resource allocation and large-scale coordination in complex systems.  With the rapid development of quantum memories, one can imagine scenarios where agents in a
  multi-agent game are equipped with pre-shared entangled resources, and communication is limited by terrain,
  latency or other constraints.  In such a situation, there may be coordinated strategies these agents could take 
  with their quantum resources that are impossible without these resources.  To date identifying such scenarios by
  hand has been challenging, but large scale empirical exploration with classical \gls{rl} that simulates the 
  use of quantum resources could provide some early insights, with the plan to eventually replace these 
  resources with quantum resources.

  \subsection*{Cross-cutting patterns}
  Three patterns recur across these domains. First, \emph{quantum data as a bridge}: in each case,
  data generated by quantum devices---whether from many-body simulation, electronic structure
  calculation, or game play---can improve classical models in ways that are provably inaccessible
  to purely classical computation. Second, \emph{classical \gls{ml} as an accelerant}: neural
  quantum states, \gls{ml}-derived potential energy surfaces, and \gls{rl}-trained game agents all
  show how classical learning makes quantum methods practical before full fault tolerance is
  achieved. Third, \emph{the virtuous cycle}: improved classical models feed back into quantum
  hardware and algorithm design, which in turn produces higher-quality quantum data for the
  next generation of classical models. These domains are where we expect the cycle to close
  first, and where the earliest demonstrations of mutual benefit are most likely to emerge.

  \section{The Path Forward}
  Classical \gls{ml} today is already providing considerable advantage to the development
  of quantum technology.  Classical \gls{ml} methods have been successfully applied to closed-loop control and calibration
  problems, \gls{qas} and circuit optimization, as well as \gls{qec}
  and the development of fault-tolerant quantum computation. Exciting new opportunities
  are also arising with the recent advances in foundation models and agentic \gls{ai} systems tailored to
  applications in science and engineering~\citep{lu2024aiscientistfullyautomated,
  gottweis2025aicoscientist, mitchener2025kosmosaiscientistautonomous,
  novikov2025alphaevolve, georgiev2025mathematicalexplorationdiscoveryscale}.
  As the power of classical \gls{ai} continues to expand, it is likely to provide even greater
  influence over the architectures of the quantum devices, and the programs we run on them.

  Each advance and improvement in quantum technology provided by \gls{ai} shortens the timeline 
  towards fault-tolerant quantum computers with considerable resources.  While \gls{nisq}
  devices today have pushed into the realm of beyond-classical performance on specialized tasks,
  we likely need to enter the era of early fault tolerance to get reliable high-quality data on
  a wide range of quantum simulation problems.  However, this regime may only be a few years away,
  and we should be prepared for what simulations and results would best empower the classical \gls{ai}
  we have developed thus far.  Empowered with data from quantum devices, we expect classical \gls{ai} will be 
  even better equipped to feed back into the design of quantum devices, fueling a virtuous cycle.

  The role of \gls{ai} within \gls{qc} development would need to expand accordingly, as these methods shift from their role as auxiliary tools towards being a core, deeply embedded component of the development stack. \Gls{ai} has a unique capability of navigating immense, highly constrained design spaces that present significant challenges for manual theoretical approaches. Whether orchestrating hardware-software co-design to discover physically realizable codes for \gls{qec}, optimizing continuous analog control pulses, or uncovering the mechanistic origins of many-body component noise, \gls{ai} can drive effective iterative refinement of quantum hardware, accelerating the timeline to robust fault-tolerant computation.

  Conversely, for \gls{qml} to deliver on its promise and robustly demonstrate quantum advantage, the field must pivot away from the attempts to load massive classical datasets into quantum memory, and towards tasks natively suited to quantum mechanics. In the near term, quantum devices will likely make their most profound mark on classical \gls{ai} by acting as physical data oracles, natively simulating classically intractable chemical and many-body interactions to provide high-fidelity training data for classical models. Research should focus on environments where quantum models are likely to demonstrate an unconditional edge over their classical counterparts: learning from states generated by quantum sensors, exploiting space-communication advantages in distributed systems, or deploying \gls{qrl} in quantum games with compact input descriptions but massive state-space complexity.

  As quantum computers grow beyond the point of simply providing training data, and enter the inference stage of computation we will be able to empirically evaluate when and where quantum components will be used.  To enable more
  researchers to work on these problems effectively, popular \gls{ml} frameworks should be more deeply integrated with quantum software and hardware backends. This should be coupled with an expansion of evaluation benchmarks, as standardized performance comparisons would enable easier model
  development. Rapid progress in classical \gls{ml} has been tightly coupled with the availability of highly prominent benchmarks, like ImageNet in case of computer vision, and there is currently no equivalent large-scale dataset of quantum states from high-fidelity simulations, with standardized predictive tasks. 
  
  At this critical juncture, there is also a distinct opportunity for \gls{ai} scientist agents to help identify novel applications of quantum computers, and derive new quantum algorithms to help solve some of the grand open challenges in computer science. These efforts should capitalize on rapid advances in agentic systems for autonomous research, bridging ideation, reasoning, simulation, and formal verification. Nevertheless, even under the assumption that the scope of \gls{qc} applications expands, there will likely always be tasks that quantum computers do only about as well as classical computers, in which case economics dictates that people will prefer to keep a mostly classical stack.  

  Despite the many identified challenges, we believe that there is a unique opportunity for \gls{ai} to
  establish itself as one of the key technologies underpinning the future of quantum computing breakthroughs. If we are able to deliver on the promise of \gls{qml}, as well as the promise of
  classical \gls{ml} in accelerating quantum computing applications, we would hope to establish a virtuous
  loop whereby the resulting mutual acceleration would help unlock deep impact across important
  scientific applications.

  \section*{Acknowledgements}
  The authors thank Adam Zalcman, Amira Abbas, Ben Jaderberg, and Ryan Babbush for detailed comments on the draft.
  \printbibliography
  \onecolumn
  \printglossary[type=\acronymtype]

\end{document}